\documentclass[twocolumn,twoside,slac_two]{revtex4}
\usepackage{graphicx}
\usepackage{fancyhdr}
\newcommand{\be}{\begin{equation}}
\newcommand{\ee}{\end{equation}}
\newcommand{\ba}{\begin{eqnarray}}
\newcommand{\ea}{\end{eqnarray}}

\begin{document}

\title{Sibyll with charm}

\author{Eun-Joo Ahn$^a$, Ralph Engel$^b$, Thomas K. Gaisser$^c$, Paolo Lipari,$^d$ Todor Stanev$^c$}
\affiliation{$^a$ Fermi National Accelerator Laboratory, Batavia, IL, 60510, USA \\
$^b$ Institut f\"ur Kernphysik, Karlsruher Institut f\"ur Technologie - Campus Nord,
76021 Karlsruhe, Germany \\
$^c$ Bartol Research Institute, University of Delaware, Newark, DE, 19716, USA \\
$^d$ INFN sezione Roma ``La Sapienze'' Dipartimento di Fisica Universit\'a di Roma I, Piazzale Aldo Moro 2, I-00185 Roma, Italy}

\begin{abstract}
The cosmic ray interaction event generator {\sc Sibyll} is widely used in extensive air shower simulations for cosmic ray and neutrino experiments. Charmed particle production has been added to the Monte Carlo with a phenomenological, non-perturbative model that properly accounts for charm production in the forward direction. As prompt decays of charm can become a significant background for neutrino detection, proper simulation of charmed particles is very important. We compare charmed meson and baryon production to accelerator data.
\end{abstract}

\maketitle

\thispagestyle{fancy}

\section{Introduction}

The cosmic ray event generator {\sc Sibyll} \cite{Fletcher:1994bd, Ahn:2009wx} is a relatively simple model that is able to reproduce many features of hadronic interactions in fixed target and collider experiments. {\sc Sibyll} is based on the dual parton model (DPM) \cite{Capella:1977me,Capella:1981xr,Capella:1992yb}, the Lund Monte Carlo algorithms \cite{Bengtsson:1987kr, Sjostrand:1987xj}, and the minijet model \cite{Gaisser:1984pg, Pancheri:1985ix, Durand:1987, Durand:1988cr}, allowing multiple hard and soft interactions. The hard and soft interactions are discriminated by an energy-dependent transverse momentum cutoff. The hard interaction cross section is calculated according to the minijet model. Diffraction dissociation is implemented as a two-channel eikonal model based on the Good-Walker model \cite{Good:1960ba}. For hadron-nucleus interactions, the interaction probability for each nucleon inside the nucleus is calculated based on the impact parameter distribution. The total interaction cross section is calculated using the Glauber scattering theory \cite{Glauber:1970jm}. For a nucleus-nucleus interaction the semi-superposition model \cite{Engel:1992vf} is used to determine the point of first interaction for the nucleons of the projectile nucleus. The fragmentation region is emphasized as appropriate for air shower simulations. Version 2.1 has been available since 1999.

The new version 2.2c has charmed quark and anti-quark added to the $u$, $d$, $s$ quarks and anti-quarks, and gluons which are present in version 2.1. Small updates have also been made to give better agreement with experimental data. Addition of charmed quark is especially relevant to air shower simulation studies of neutrino detection: the short-lived charmed particles create ``prompt'' muons and neutrinos in the atmosphere with a harder spectrum and isotropic distribution, whose contributions are searched for in the atmospheric muon and neutrino spectrum. A proper representation of charmed particles is thus very important for simulations involving neutrino experiments.

In the following sections, we describe how the charmed quark is added to {\sc Sibyll}, and compare the charmed meson and baryon predictions with experimental data. Hereafter, a ``$q$ quark'' includes both the quark and anti-quark unless specified.

\section{Charm inclusion}

The following charmed particles have been added: mesons $D$, $D^\star$ families, $\eta_c$, and $J/\psi$; and baryons $\Sigma_c,~ \Sigma_c^\star,~ \Xi_c,~ \Xi_c^\star$ families, $\Lambda_c^+$, and $\Omega_c^0$.

Charmed particles are formed throughout the whole fragmentation process. The basic assumption is that $c$ quarks form in all processes where $s$ quarks form, but with lower frequency.  This is realised in the Monte Carlo by replacing production of an $s\bar{s}$ pair by a $c\bar{c}$ pair with a frequency $P_c/P_s = 0.004$.  The numerical value is adjusted for a best fit to the data.  For comparison, $P_s/P_d = P_{us}/P_{ud} = 0.3$.  Valence $s$-quarks (for example in kaon projectiles) are excluded from this substitution process. This procedure ensures that all charm production coming from hard and soft processes are covered. The cross section is controlled by the branching ratio.

In a collision, the leading particle carries a valance quark or diquark of the projectile, in contrast to the non-leading ones. Leading particles therefore are more abundantly produced in the forward region, i.e. large Feynman $x_F$, resulting in an asymmetry between leading and non-leading particles. This leading quark effect is well established in charm hadroproduction. For example, the quark composition of a proton projectile favours $\Lambda_c^+$ over $\bar{\Lambda}_c^-$ as leading particles. Charmed particles formed via fragmentation allows the leading quark effect to be observed \cite{Garcia:2001xj} .

The momenta assigned to the charmed particles differs from non-charmed particles to give a harder spectrum. While the non-charmed particles follow the Lund fragmentation function, the heavier charmed particle fragments according to the Peterson/SLAC fragmentation function \cite{Peterson:1982ak}
\be
\displaystyle
f(z) ~\propto~ \left[ z \left( 1 \,-\, \frac{1}{z} \,-\, \frac{\epsilon_Q}{1-z} \right)^2 \right]^{-1} \ ,
\ee
where $z$ is the fraction of the new particle energy with respect to the parent quark or diquark, and the free parameter $\epsilon_Q$, which is inversely proportional to the square of the heavy quark mass, has been set to 2. The primordial $p_T$ assigned to the newly formed $c$ quark or diquark pairs during fragmentation has a Gaussian distribution with an energy-dependent mean value
\be
\langle p_T \rangle ~=~ \left[ p_0 \,+\, 0.08 \log_{10} \left( \frac{\sqrt{s}}{30 \, \rm{GeV}} \right)\right] {\rm GeV/c} \  ,
\label{eq:softpt}
\ee
where $p_0 = 1.5$ GeV/c and 1.0 GeV/c for baryons and mesons, respectively. This setting results in a higher  $\langle p_T \rangle$ for charmed particles, compared to other quarks and diquarks where the $\langle p_T \rangle$ ranges from 0.3 to 0.6 GeV/c.

Charmed particles are produced in the interactions of high-energy primary cosmic rays. The quick decay of these charmed particles form the ``prompt'' muons and neutrinos in the atmosphere. These prompt leptons are expected to have an isotropic angular distribution and a harder spectrum than particles produced from ordinary decay of pions and kaons. Because of their relatively hard spectrum and isotropy, prompt neutrinos are a significant background in the search for astrophysical neutrino signals above 100 TeV.

Additional minor modifications with respect to {\sc Sibyll} 2.1 have been made in version 2.2c.  These include improvements in treatment of diffraction, increasing the $s$ quark fraction and fixing a bug in the energy dependence of the cutoff of $p_T$. These changes give a higher multiplicity and a smoother transition to diffraction in the forward direction.

\section{Comparison with data}

Most information on differential behaviour is obtained from fixed target experiments, with hadron-hadron or hadron-nuclei collisions. Distributions of Feynman $x_F$ and transverse momenta $p_T^2$ from the LEBC bubble chambers, E769, and SELEX experiments are used to compare the charmed meson and baryon behaviour with {\sc Sibyll} 2.2c.

\begin{figure}[h]
\includegraphics[width=80mm]{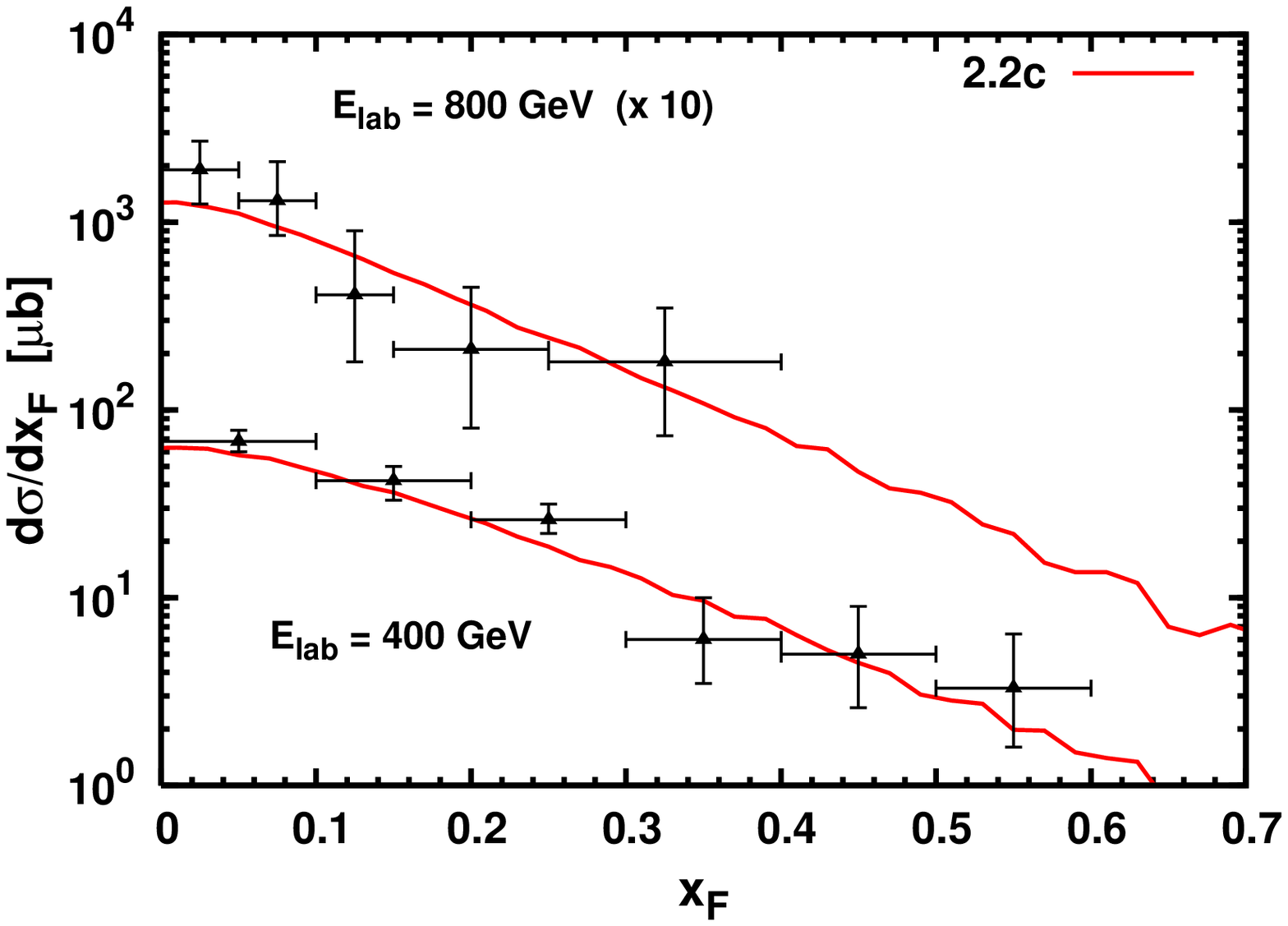}
\includegraphics[width=80mm]{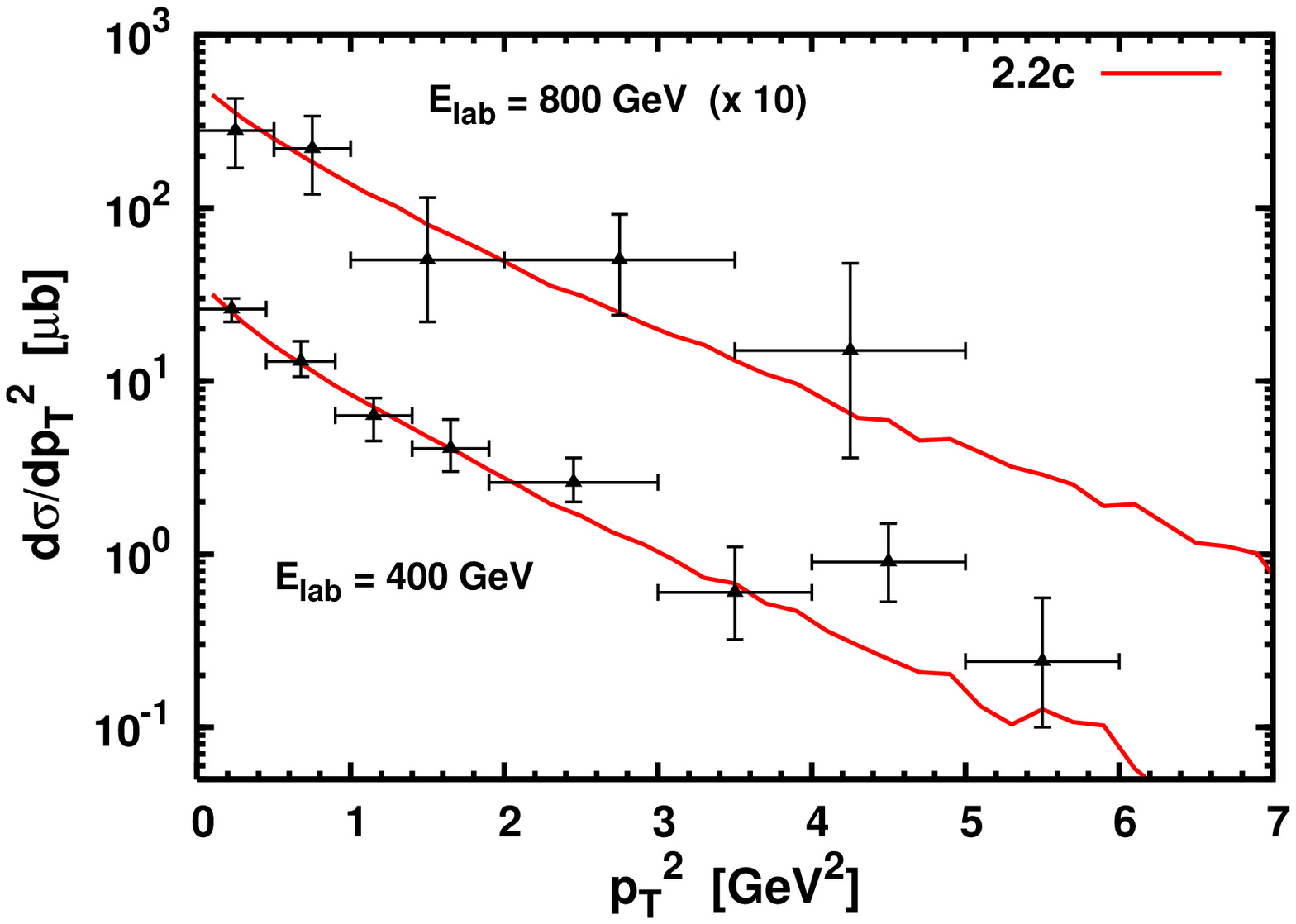}
\caption{Differential cross section distributions of $x_F$ (upper panel) and $p_T^2$ (lower panel)  for $p$-$p$ collisions producing $D$s from {\sc Sibyll 2.2c} plotted against data from LEBC-EHS ($E_{lab} = 400$ GeV) and LEBC-MPS ($E_{lab} = 800$ GeV). For clarity purpose, the simulation results and data for LEBC-MPS has been multiplied by factor 10. }
\label{fig:lebc}
\end{figure}

The two bubble chamber experiments LEBC-EHS \cite{AguilarBenitez:1988sb} at $E_{lab} = 400$ GeV and LEBC-MPS \cite{Ammar:1988ta} at $E_{lab} = 800$ GeV used $p$-$p$ interactions to search for production of all $D$ particles. Figure \ref{fig:lebc} compares the $D$ production of {\sc Sibyll} 2.2c to $x_F$ and $p_T^2$ distributions of the differential cross sections. All the comparisons show good agreement with data.

\begin{figure*}[h]
\hspace*{-5mm} \includegraphics[width=60mm]{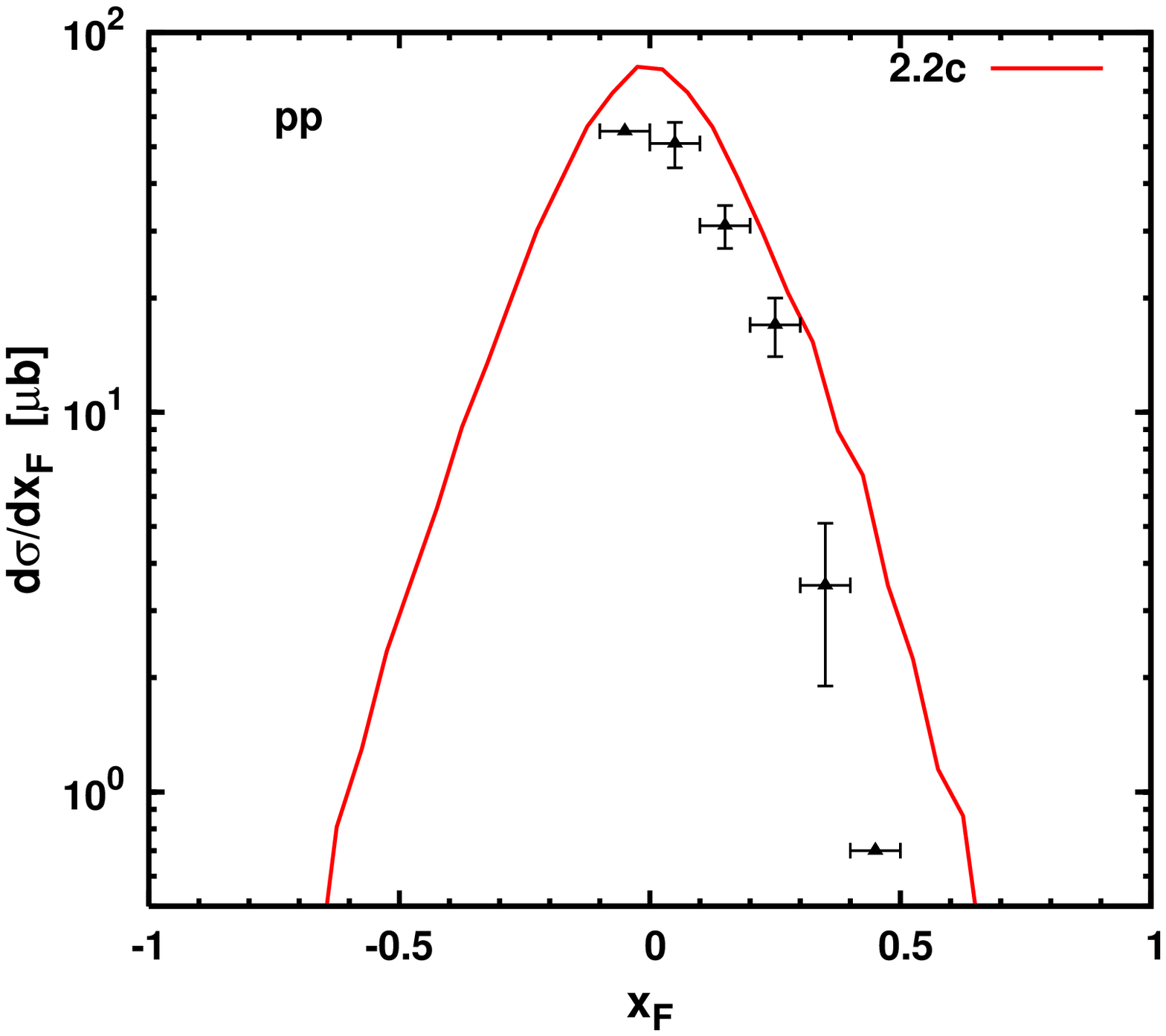}  
\hspace*{-5mm} \includegraphics[width=60mm]{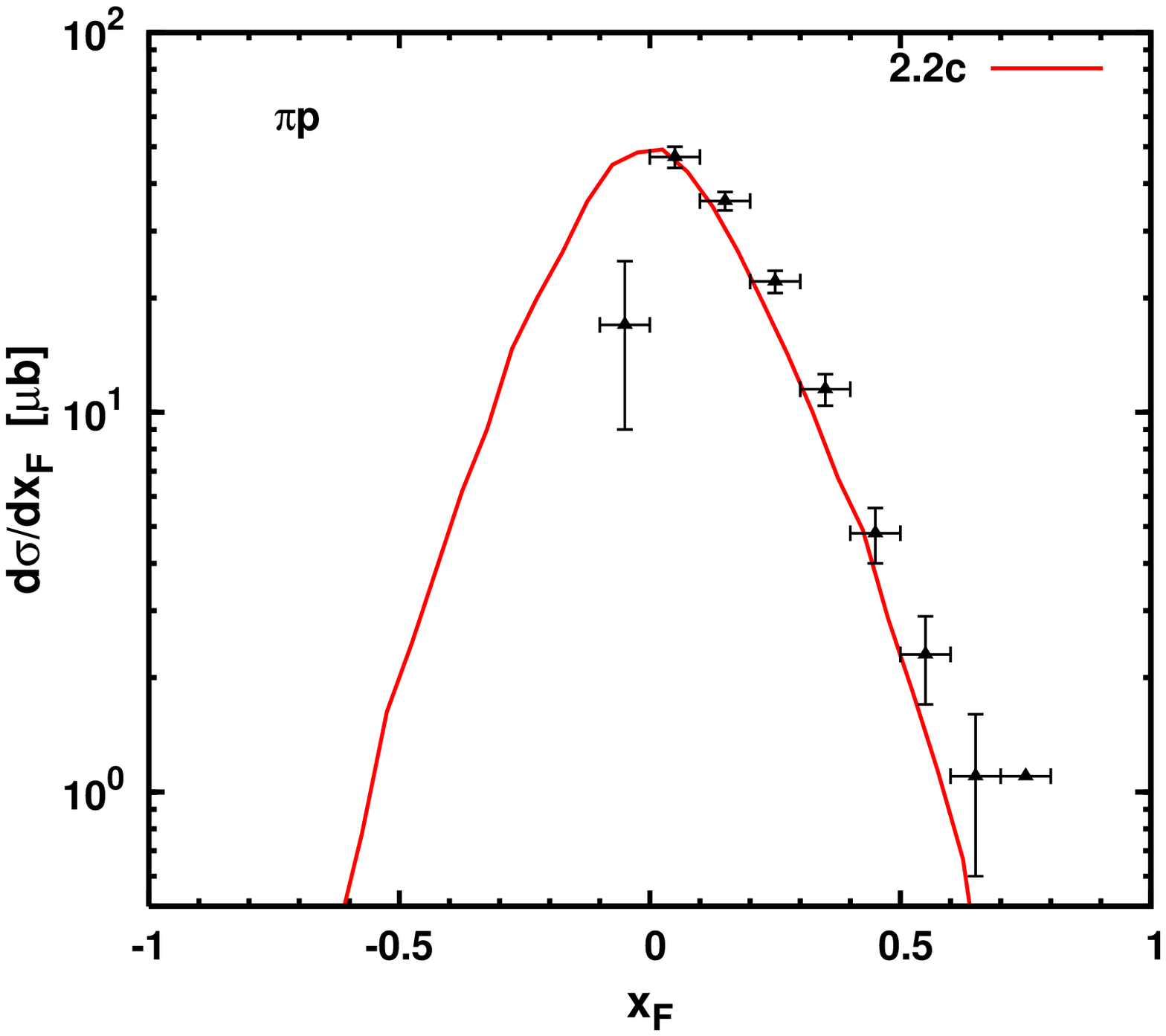} 
\hspace*{-5mm} \includegraphics[width=60mm]{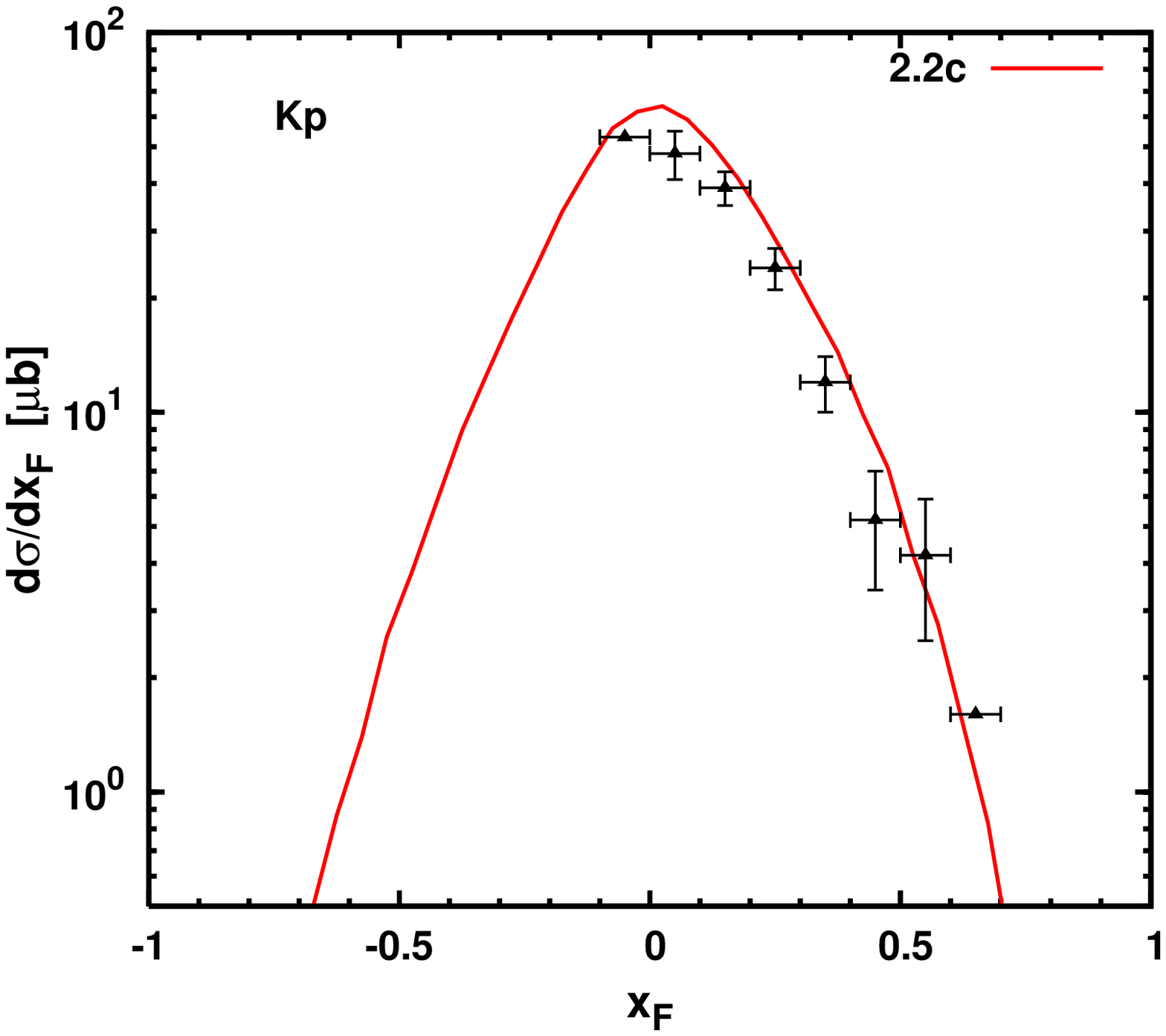} \\
\vspace*{-2mm}
\hspace*{-5mm} \includegraphics[width=60mm]{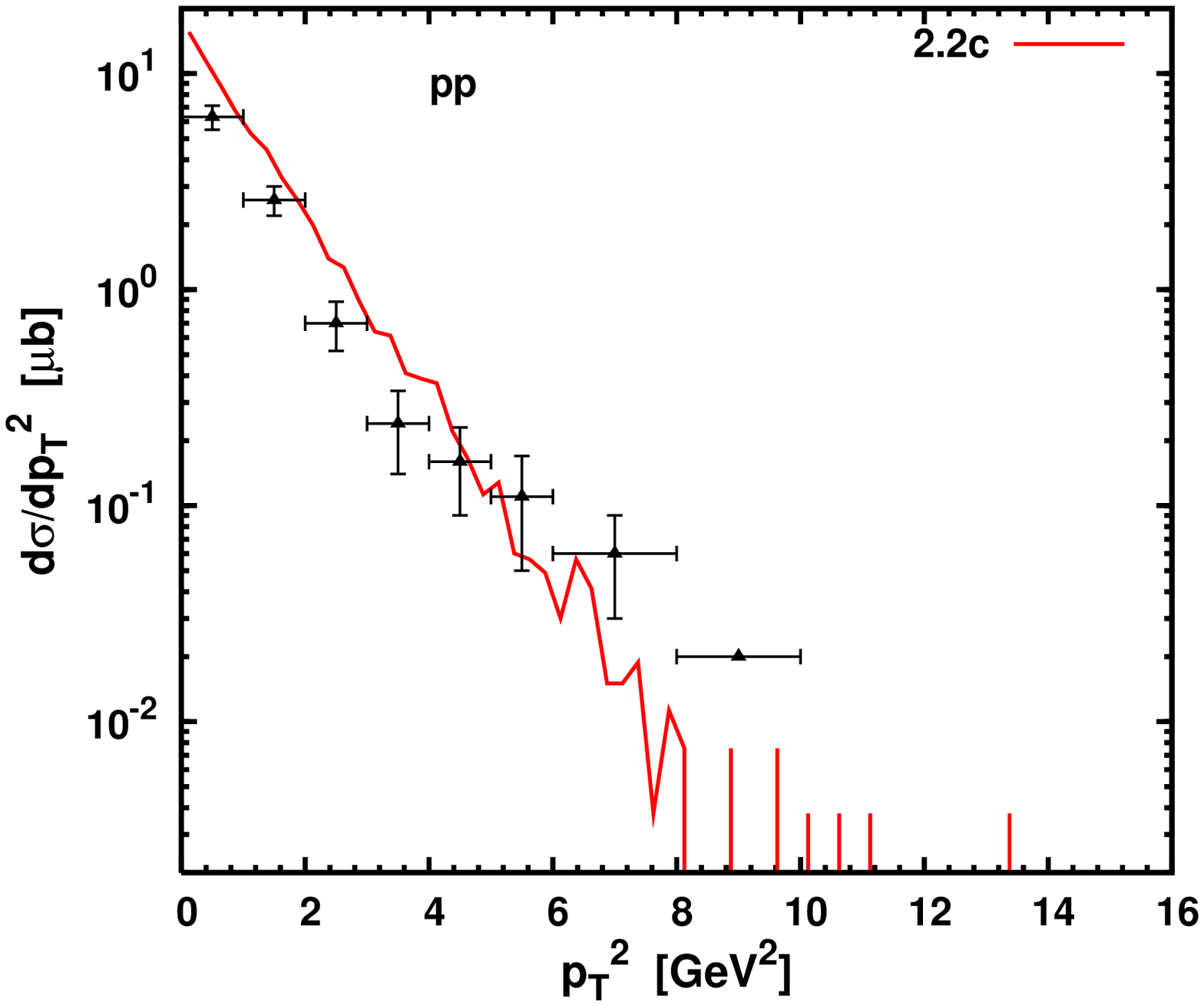} 
\hspace*{-5mm} \includegraphics[width=60mm]{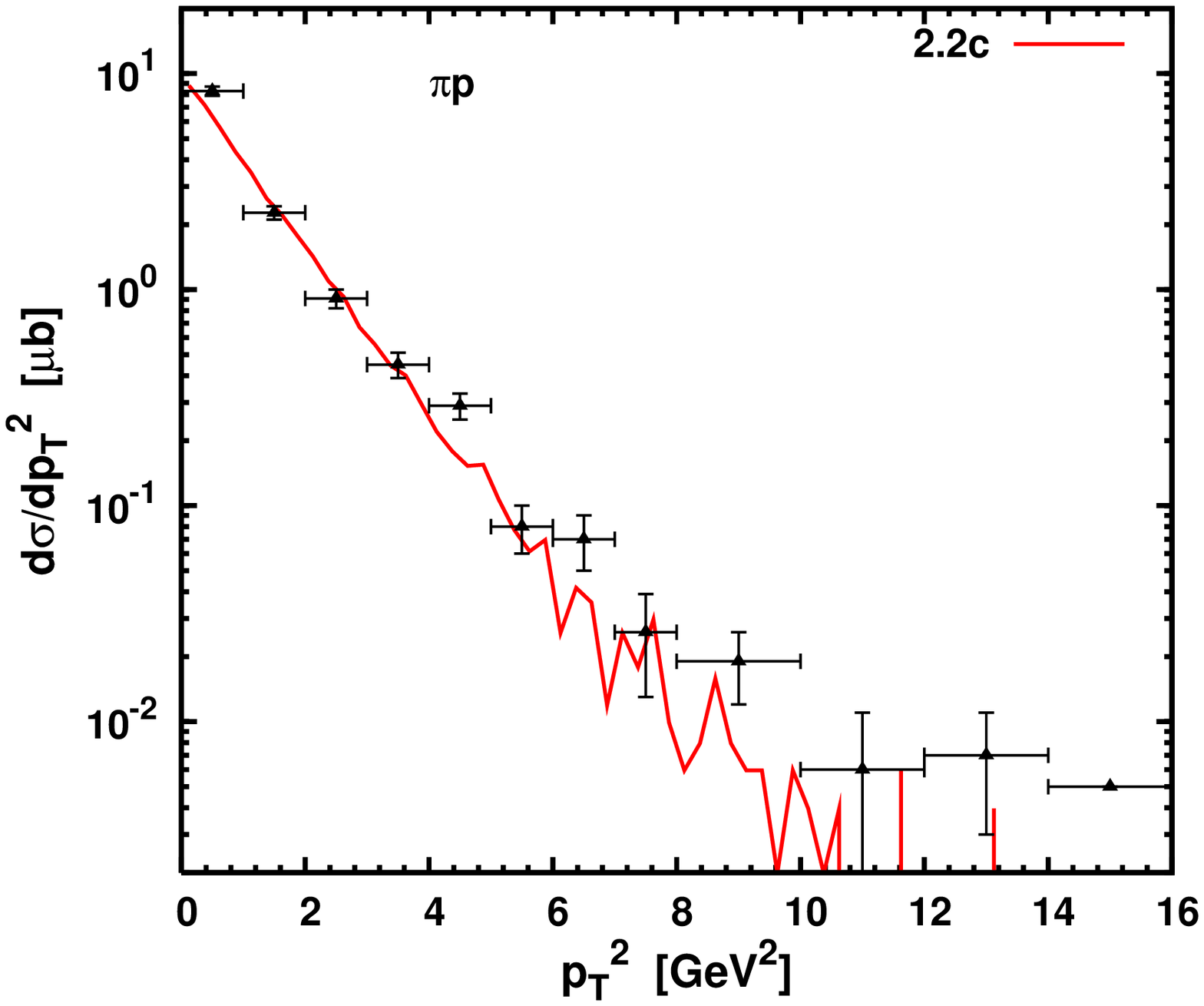} 
\hspace*{-5mm} \includegraphics[width=60mm]{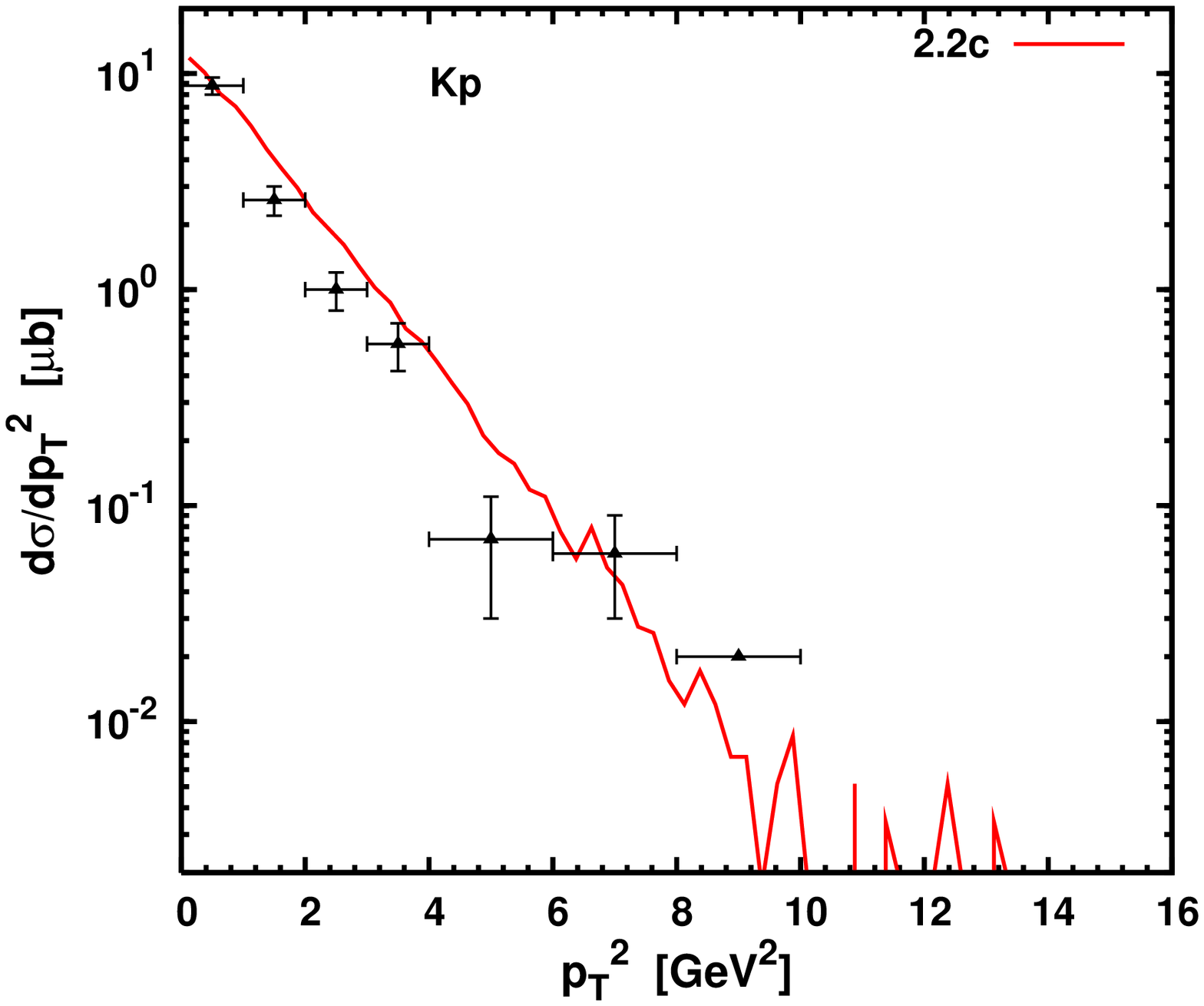}
\caption{Differential cross section distributions of $x_F$ (upper panels) and $p_T^2$ (lower panels) for $D$ production per target nucleon for {\sc Sibyll 2.2c}. The interactions $p$-$p$ (left panels), $\pi$-$p$ (centre panels), and $K$-$p$ (right panels) are compared with the data from E769. }
\label{fig:e769}
\end{figure*}

The E769 experiment \cite{Alves:1996qz} used $p,~ \pi^\pm,~ K^\pm$ projectiles on a variety of nuclear targets to obtain the $x_F$ and $p_T^2$ behaviour of the $D$ mesons, where the differential cross sections are given per target nucleon. Within the perturbative QCD treatment of heavy particle production, the incident hadron sees the nucleus as individual nucleons.  As a consequence, the production cross section is expected to scale with the nuclear mass $A$. Such behaviour has been verified by the E769 and WA82 \cite{Adamovich:1992fx} experiments. Nuclear scaling of the target has been upheld in the simulations, and a proton target has been used to compare with the data. Figure \ref{fig:e769} shows the $D$ production of {\sc Sibyll} 2.2c plotted against the E769 data. The overall agreement with data is good, although the simulated $p_T^2$ spectra are somewhat softer than data.

\begin{figure*}[h]
\includegraphics[width=66mm]{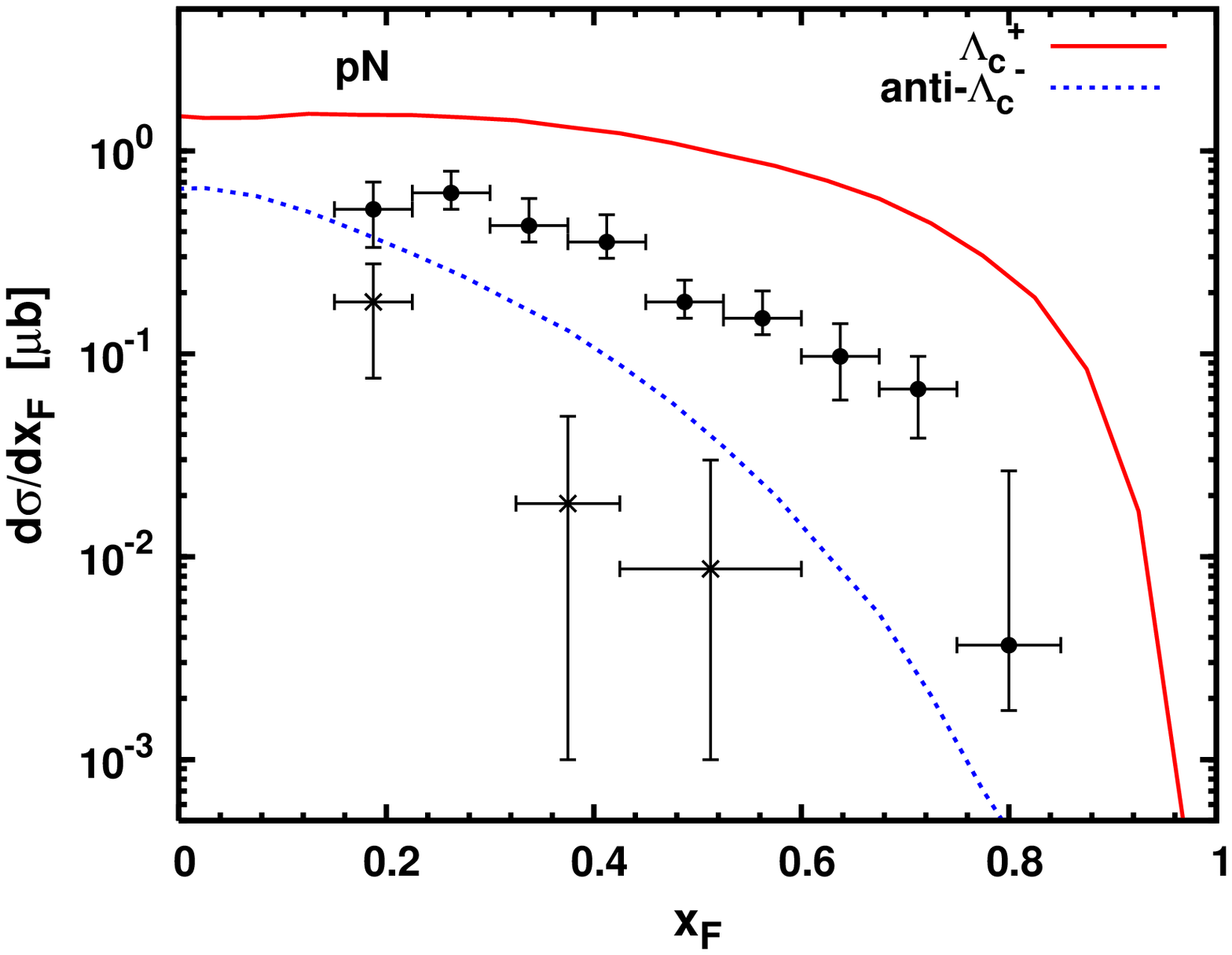}
\includegraphics[width=66mm]{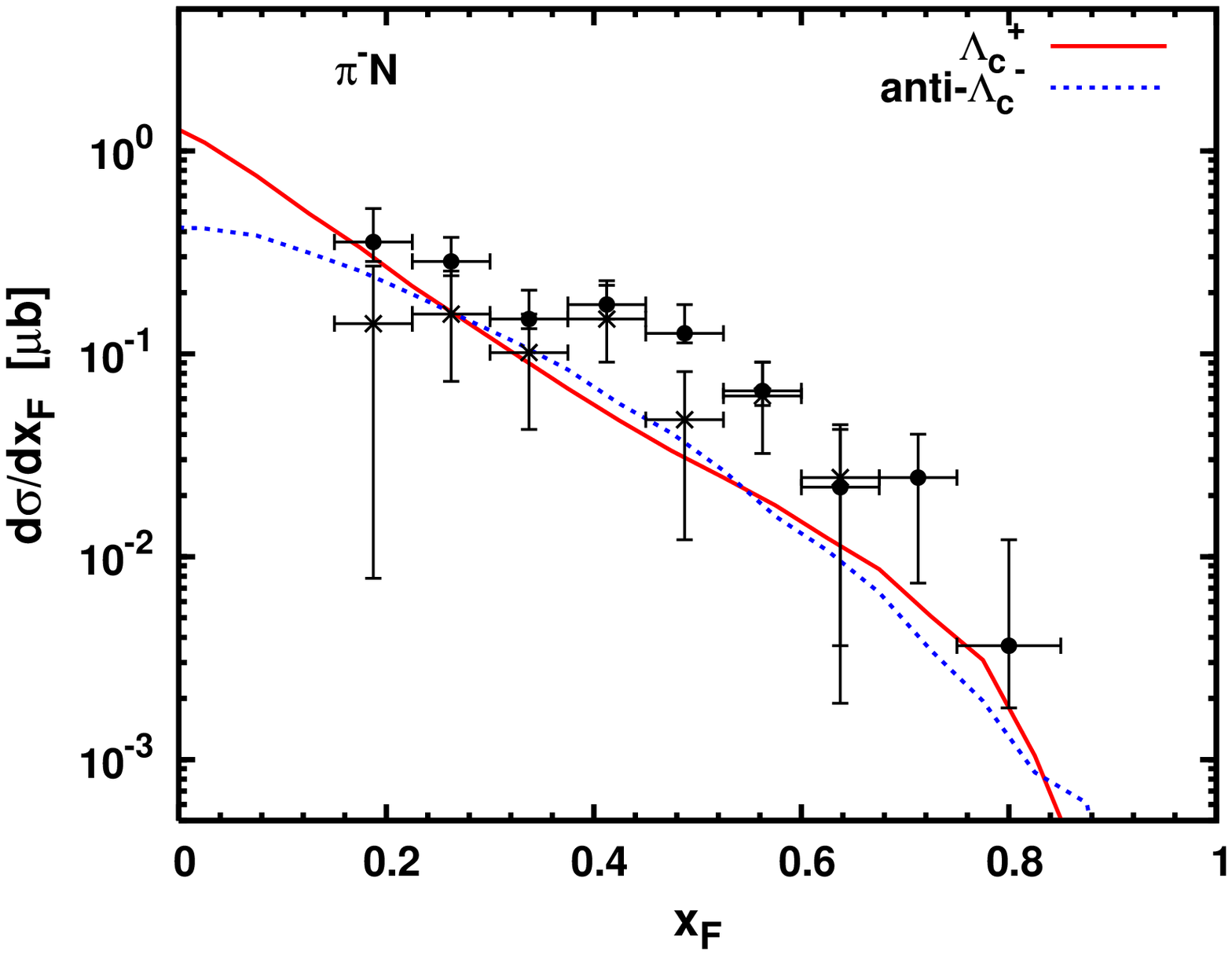} \\
\vspace*{-2mm}
\includegraphics[width=66mm]{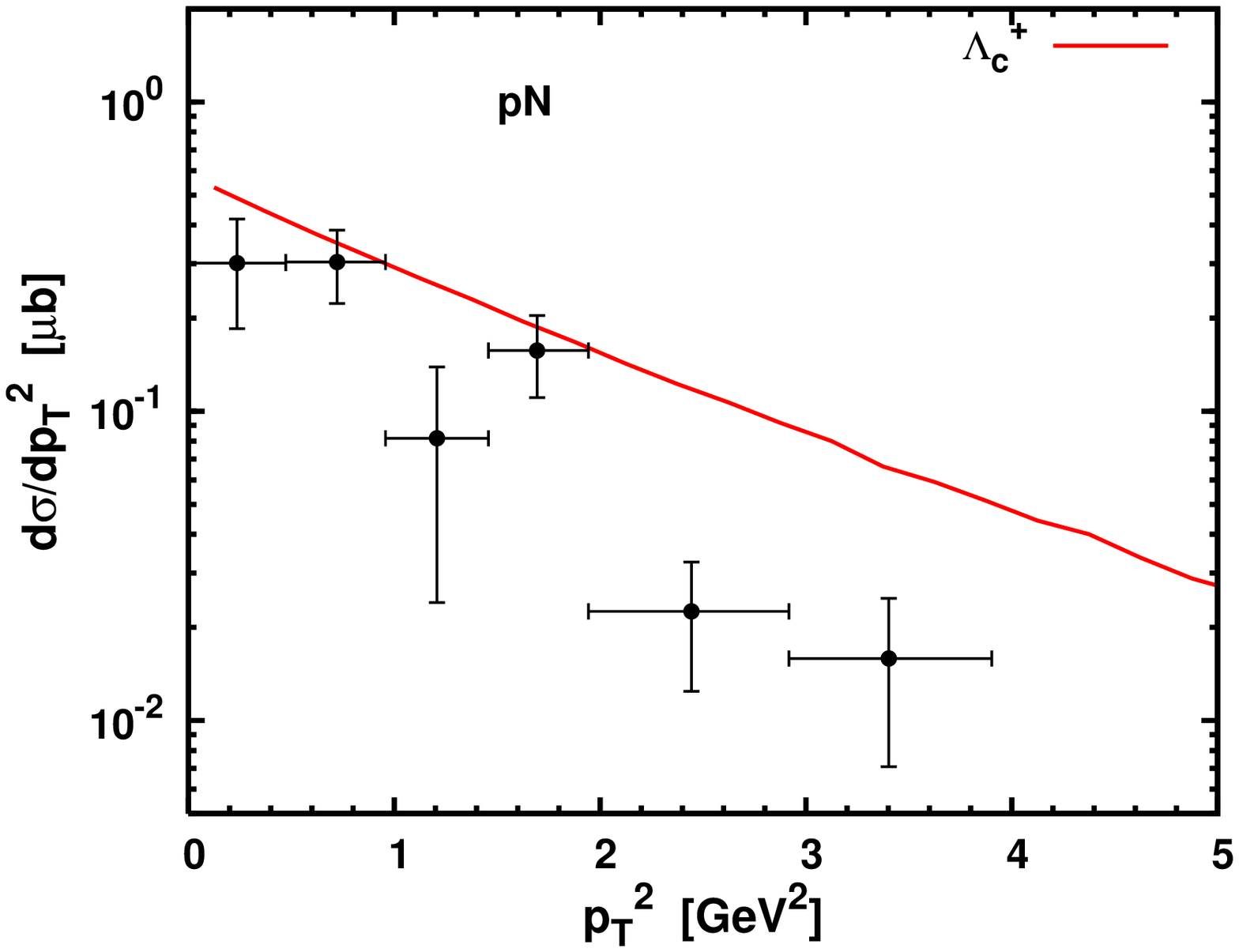}
\includegraphics[width=66mm]{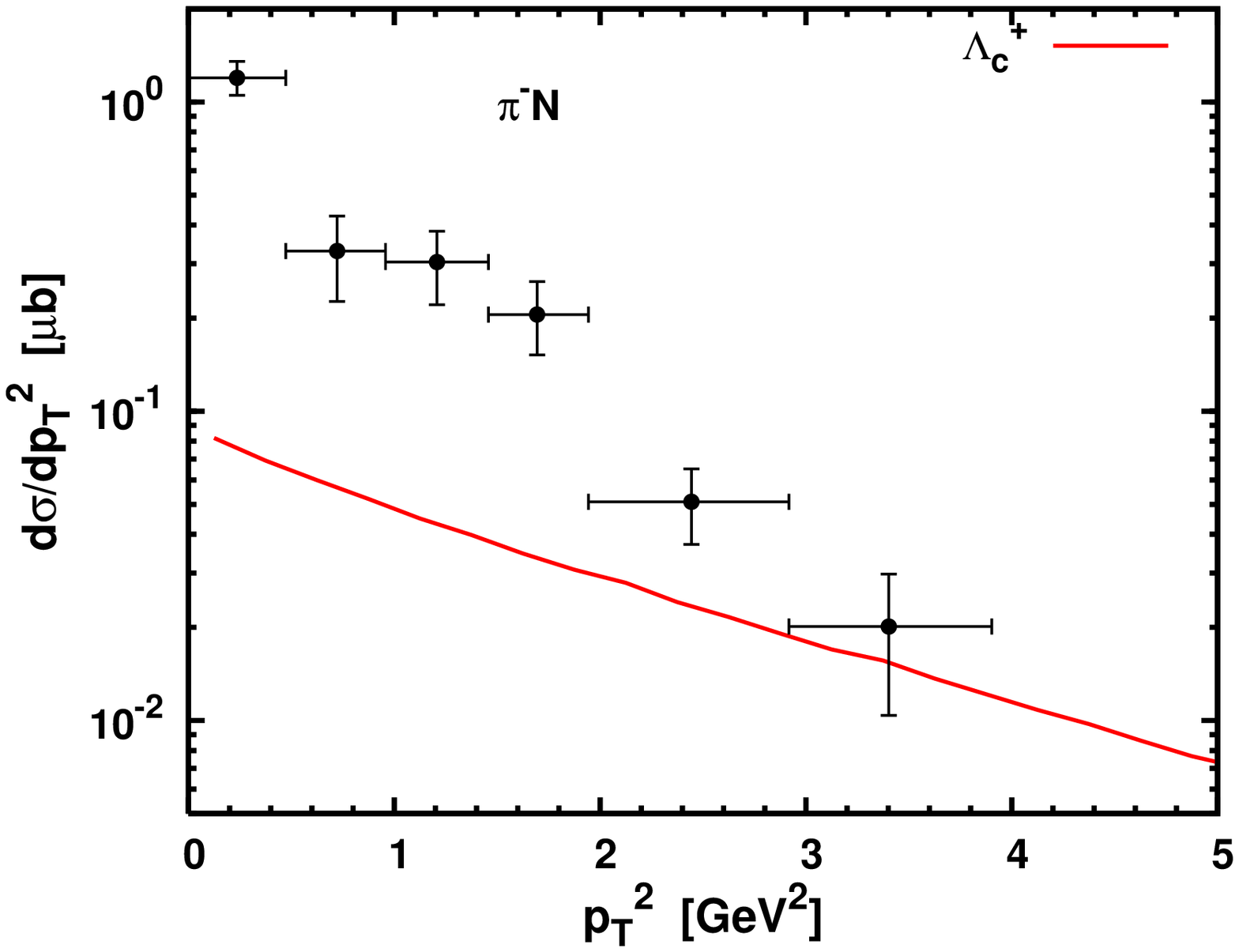}
\vspace*{-3mm}
\caption{Differential cross section distribution of $x_F$ (upper panels) and $p_T^2$ (lower panels) with $p$ (left panels) and $\pi^-$ (right panels) projectiles. The $\Lambda_c^+$  ( $\bar{\Lambda}_c^-$) production from {\sc Sibyll} 2.2c shown in red solid (blue dotted) lines are compared with data from SELEX.}
\label{fig:selex}
\end{figure*}

The SELEX experiment \cite{Garcia:2001xj} used $\pi^-$ and $\Sigma^-$ projectiles at $E_{lab} = 615$ GeV and $p$ projectile at $E_{lab} = 540$ GeV, on a variety of nuclear targets to obtain the distribution of charmed baryon. The leading quark effect in charm hadroproduction of baryons can be tested, as the lack of anti-quarks in the baryon beams $\Sigma^-$ and $p$ favours quark over anti-quark and an asymmetry between $\Lambda_c^+$ and $\bar{\Lambda}_c^-$ production is expected. SELEX observed such an asymmetry with the baryon beams  and reported a relatively flat and hard $x_F$ distribution, fitting to the parameterisation of $(1-x_F)^{2.5}$. As only the number of events of the distribution of $x_F$ and $p_T^2$ are given with no absolute cross section measurement, the normalisation suggested by Ref.  \cite{Berghaus:2007hp} has been adopted, where the SELEX data are multiplied by $ 7 \mu$b/nucleon. In the simulation, $\pi^-$ and $p$ beams are used against nitrogen nuclei target. Comparison of {\sc Sibyll} 2.2c with the data is shown in Fig.~\ref{fig:selex}. For the proton beam, the shape of both $x_F$ and $p_T^2$ for $\Lambda_c^+$ distribution are in agreement though they appear to be over-produced using the current normalisation. For the $\pi^-$ beam, the $\Lambda_c^+$ are underproduced with a slightly harder $p_T^2$ distribution. Due to the lack of cross section information, we are less concerned with the overall scaling. The simulation agrees with the general trend of the data.

\section{Summary}

A new version of {\sc Sibyll} 2.2c has been presented. This version is updated from the existing version 2.1 to include production of charmed quarks.  In addition, several minor changes have been made to give a better realisation of the underlying model.  The method of substituting a small fraction of $s\bar{s}$ production by $c\bar{c}$ production covers the full phase space for charm production. There is good agreement with charmed meson and baryon data. The model reproduces the observed asymmetry between leading $\Lambda_c^+$ and $\bar{\Lambda}_c^-$. Version 2.2c will be especially useful in conducting simulations for astrophysical neutrino detection experiments such as Baikal \cite{Aynutdinov:2005dq}, ANTARES \cite{Aguilar:2006pd}, and IceCube \cite{Achterberg:2006md, Karle:2010xx}.

\section{Acknowledgements}
Work on this project at the University of Delaware is supported in part by a grant from the Office of Science of the U.S. Department of Energy, DE-FG02-91ER40626. Fermilab is operated by Fermi Research Alliance, LLC under Contract No. DE-AC02-07CH11359 with the United States Department of Energy.

\bibliography{sibyll-22c.bib}

\end{document}